\documentclass[12pt]{iopart}

\usepackage{amsthm}
\usepackage{amssymb}

\usepackage{color}

\usepackage[figuresright]{rotating}
\usepackage{amsfonts,amssymb,cite,graphicx,color}
\newfont{\fp}{msbm10 at 11pt}  
\def\E{\mbox{\fp E}}

\newtheorem{definition}{Definition}

\newtheorem{prop}{Proposition}

\usepackage{amsfonts}

\usepackage{leftidx}

\usepackage{color, colortbl, framed}

\usepackage{leftidx}
\usepackage{graphicx}
\usepackage{tikz}
\usetikzlibrary{shapes.geometric, arrows}

\tikzstyle{startstop} = [rectangle, rounded corners, minimum width=3cm, minimum height=1cm,text centered, draw=black, fill=red!30]
\tikzstyle{process} = [rectangle, minimum width=3cm, minimum height=1cm, text centered, draw=black, fill=orange!30]
\tikzstyle{arrow} = [->,>=stealth]
\usepackage{placeins}
\usepackage{color}

\usepackage{mathptmx}       
\usepackage{helvet}         
\usepackage{courier}        
\usepackage{type1cm}        
\usepackage{makeidx}         
\usepackage{graphicx}        
                \usepackage[T1]{fontenc}
\usepackage{amsfonts}

\usepackage{leftidx}

\expandafter\let\csname equation*\endcsname\relax
\expandafter\let\csname endequation*\endcsname\relax
\usepackage{amsmath}

\usepackage{amssymb,epsfig,graphicx}

\usepackage{graphicx}             
\usepackage{multicol}        
\usepackage[bottom]{footmisc}

\def\E{\mbox{\fp E}}

\begin{document}
\newcommand{\Fracderiv}[2]{\leftidx{_0}{\mathcal{D}_{#1}^{#2}}}

\title{Fractional Brownian motion time-changed by gamma and inverse gamma process
}

\author{Arun Kumar, Agnieszka Wy{\l}oma{\'n}ska, Rafa{\l} Po{\l}ocza{\'n}ski, S. Sundar }


\address{
A. Kumar\\
Indian Statistical Institute Chennai Centre, MGR Knowledge city, CIT Campus, Taramani, Chennai, Tamilnadu, India - 600113\\
A. Wy{\l}oma{\'n}ska, R. Po{\l}ocza{\'n}ski\\
             Faculty of Pure and Applied Mathematics, Hugo Steinhaus Center, Wroc{\l}aw University of Science and  Technology, Wybrze\.ze Wyspia{\'n}skiego 27, 50-370 Wroc{\l}aw, Poland \\
S. Sundar\\
Department of Mathematics, Indian Institute of Technology Madras, Chennai, Tamilnadu, India - 600036\\

         }

\date{Received: date / Accepted: date}

\ead{arunk@isichennai.res.in, agnieszka.wylomanska@pwr.edu.pl, rafal.poloczanski@pwr.edu.pl, slnt@iitm.ac.in}

\begin{abstract}
Many real time-series exhibit behavior adequate to long range dependent data. Additionally very often these time-series have constant time periods and also have characteristics similar to Gaussian processes although they are not Gaussian. Therefore there is need to consider new classes of systems to model these kind of empirical behavior.  Motivated by this fact in this paper we analyze two processes which exhibit long range dependence property and have additional interesting characteristics which may be observed in real phenomena. Both of them are constructed as the superposition of fractional Brownian motion (FBM) and other process. In the first case the internal process, which plays role of the time, is the gamma process while in the second case the internal process is its inverse. 
We present in detail their main properties paying main attention to the long range dependence property. Moreover, we show how to simulate these processes and estimate their parameters.  We propose to use a novel method based on rescaled modified cumulative distribution function for estimation of parameters of the second considered process. This method is very useful in description of rounded data, like waiting times of subordinated processes delayed by inverse subordinators. By using the Monte Carlo method we show the effectiveness of proposed estimation procedures. 

\end{abstract}
\textbf{Keywors:} subordination, gamma process, inverse gamma process, simulation, estimation 



\submitto{Journal of Statistical Mechanics: Theory and Experiment}
\maketitle

\section{Introduction}\label{intro}
In recent years time-changed stochastic processes (TCSP) are getting increasing attention. Note that a TCSP are obtained by changing the time  of a stochastic process by some other process which is generally a non-decreasing L\'evy process called also subordinator. We mention, a L\'evy process has stationary and independent increments with c\'adl\'ag sample paths. TCSP are used  for example in finance to provide stochastic volatility models \cite{Barn-Nic-Shep}. Further, they are also used in statistical physics to model anomalous diffusion phenomena \cite{wyl1}.  TCSP are convenient way to develop a model where it is desired to retain some properties of the outside process (called also external process) and at the same time it is required to change other characteristics. TCSP are also called some time subordinated stochastic processes. The idea of subordination was introduced in 1949 by Bochner \cite{boch} and expounded in his book in \cite{boch2}. The theory of subordinated processes is also explored in details in \cite{sato}. The subordinated processes were studied in many areas of interest, for example in finance \cite{fin1,fin2,fin3,fin4,ding-granger,pagan}, physics \cite{fiz1,fiz2,fiz3,fiz4}, ecology \cite{ec}, hydrology \cite{dok-opp-taqqu} and biology \cite{bi}.\\
The long-range dependence (LRD) phenomena, called also long memory, first was introduced by Mandelbrot and Wallis in 1968 \cite{k1}. Since that time the models and processes which exhibit long memory property have found many practical applications. Moreover, number of research papers related to this phenomena are increasing rapidly. One of the most classical model used  to describe data sets with LRD is the FBM \cite{NessMandelbrot} which is closely related to fractional Langevin equation motion \cite{Metzler_fractional} and  is a generalization of the classical Brownian motion. It is worth to mention, the FBM is also one of the main model used for description of data with anomalous diffusion behavior.  One can observe LRD in many real data, for example, economic and financial time-series \cite{dok-opp-taqqu,k2}, physics and natural sciences data \cite{k4,k5}, as well as hydrology, meteorology and geophysics time series \cite{NessMandelbrot,k7,k8,k9}. In \cite{w47} one can find additional interesting examples of long-range dependent time series, like single particle tracking dynamics in molecular biology, electromagnetic field data or high solar flare activity. The LRD relates to the rate of decay of statistical dependence of two points with increasing lag or spatial distance between the points. A phenomenon is usually considered to have LRD if the dependence decays more slowly than an exponential decay, typically a power-like decay. What should be mentioned, L\'evy processes can not model data that exhibits LRD and most of LRD data come very often from non-Gaussian distributions.\\ 
To keep these requirements in mind, we study FBM time-changed by gamma and inverse gamma process. By changing the time of FBM with gamma and inverse gamma stochastic processes, our aim is to retain the LRD property of FBM and to introduce heavy-tailedness and time varying volatility. Note that gamma process is a L\'evy process with increments having gamma distribution while the inverse gamma process is the first-exit time of gamma process. It is worth to mention here that there is an inverse gamma L\'evy process proposed in literature (see e.g. \cite{heyde-leonenko}) however it is different from  definition given in this paper. The first analyzed process, known also as the fractional Laplace motion (FLM)\cite{meer}, can be useful for modeling data having LRD property and non-Gaussian behavior. The density of FLM is not Gaussian, however, it  exhibits many properties of Gaussian-based processes, like existence of moments. The second considered process can be used for data with LRD and visible constant time periods characteristic. \\
Like FBM, the processes time-changed by inverse subordinators are considered as models adequate for anomalous diffusion phenomena. In the domain of anomalous diffusion the typical approach is based on continuous time random walk (CTRW) \cite{ar30,ar31}, and the subordinated L\'evy process can be considered as a scaling limit of CTRW \cite{ar32}. The main point in framework of CTRW and subordination technique is distribution of waiting-times, which correspond to observed constant time periods \cite{ar33}. In recent years number of papers devoted to this issue have grown rapidly. We only mention, for instance an inverse $\alpha-$stable subordinator was considered in \cite{ar32,ar34,ar35}, the inverse tempered stable subordinator was examined in \cite{ar33,ar36,ar38}, while the inverse gamma process as a time-change was mentioned in \cite{ar39}. The general case inverse subordinators based on infinite divisible processes were explored for example in \cite{ar40,ar41,ar42}. \\
In this paper we present the main properties of gamma and inverse gamma processes, like tail behavior and structure of dependence. We also proved the main characteristics of FBM time-changed by these processes. Although the FLM was examined in \cite{meer}, here we mention the properties of it in order to compare with the appropriate characteristics of FBM delayed by inverse gamma subordinator.  We concentrate on the density functions of both processes, asymptotic behavior of moments and the covariance structures to establish the LRD property for both process under study. Moreover, we indicate that both systems can be classified as anomalous diffusive.\\
In this paper we also present the simulation procedures for both considered time-changed processes. The procedures are based on the fact that the analyzed systems are superpositions of two independent processes. Further, we also propose parameters estimation schemes for both processes. Note that estimation of FLM parameters are not discussed in \cite{meer}. Estimation techniques are different in both cases. For the FLM we use the fact that the mean square displacement (MSD) has the same asymptotic behavior as for the FBM, therefore, we estimate the Hurst exponent by using this measure. The parameter corresponding to gamma process are estimated using asymptotic behavior of the tail of FLM. For the FBM time-changed by inverse gamma process we use visible constant time periods observed in the trajectories of considered process. The waiting times observable in the data constitute sample of independent random variables with gamma distribution. In order to estimate the parameter of this distribution we use here a new method based on the rescaled modified cumulative distribution function introduced in \cite{rafal} which is very effective for rounded data like waiting times. The effectiveness of estimation procedures is checked by using simulated trajectories of considered processes. The obtained results indicate those methods can be used for real data which has similar properties as theoretical models. \\
The rest of the paper is organized as follows: in section \ref{def} we introduce two considered processes, namely time-changed FBM by gamma and inverse gamma processes. Next, in section \ref{theory} we present the main properties of gamma process, inverse gamma process and both time-changed processes. In section \ref{sim} we explain how to simulate the trajectories of both time-changed processes while in section \ref{est} we provide step by step procedures of estimation of their parameters. Last section contains conclusions.

\section{FBM delayed by gamma and inverse gamma process}\label{def}
\begin{figure}[h!]
\begin{center}
\includegraphics[width=0.6\textwidth]{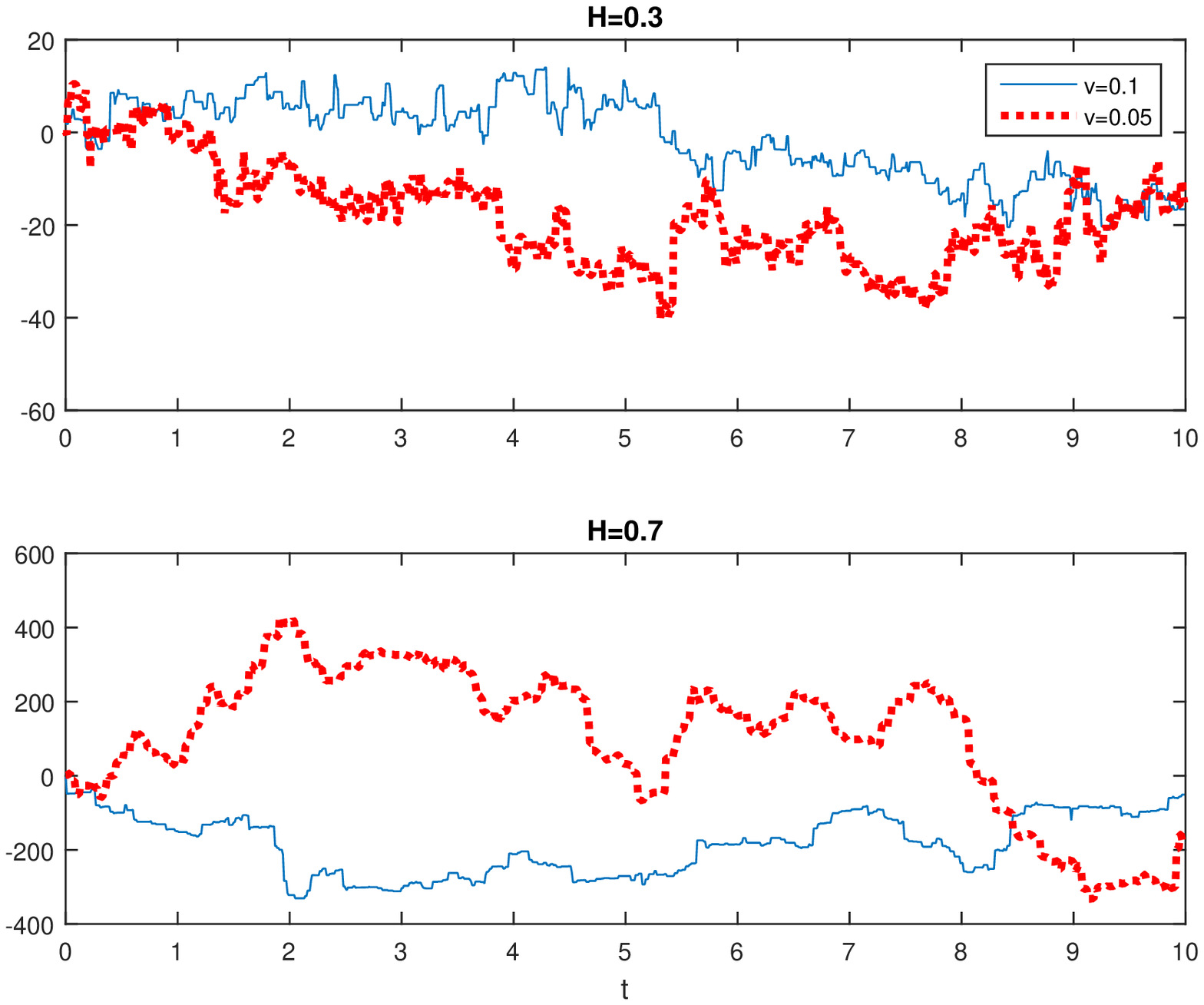}
\caption{The exemplary trajectories of the process $Y_1(t)$ for $H=0.3$ (top panel) and $H=0.7$ (bottom panel). The blue solid lines correspond to parameter $\nu=0.1$ while the red dotted lines to $\nu=0.05$. The process was simulated on the interval $[0,10]$. }
\label{fig1}  
\end{center}
\end{figure}

\begin{figure}[h!]
\begin{center}
\includegraphics[width=0.6\textwidth]{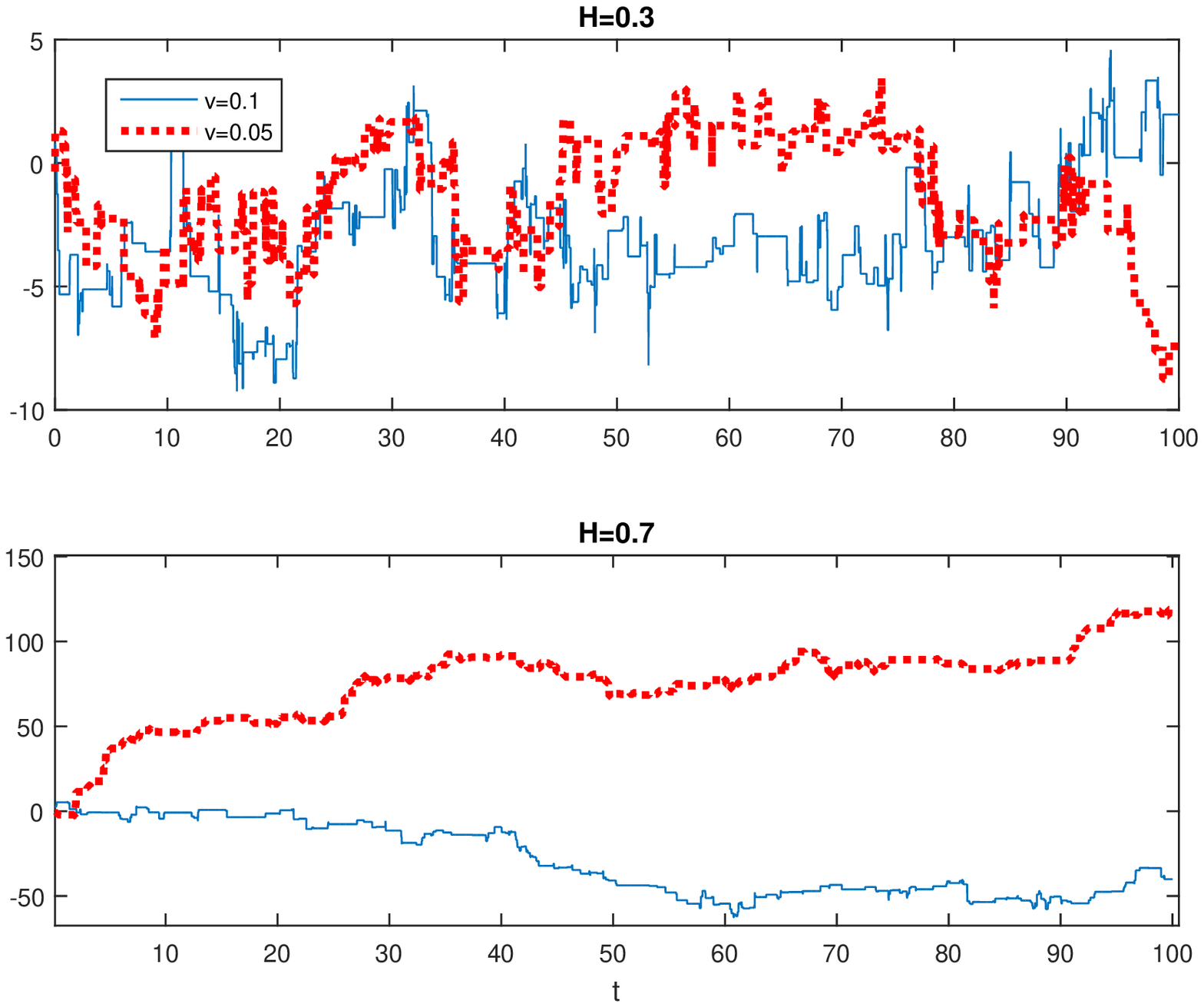}
\caption{The exemplary trajectories of the process $Y_2(t)$ for $H=0.3$ (top panel) and $H=0.7$ (bottom panel). The blue solid lines correspond to parameter $\nu=0.1$ while the red dotted lines to $\nu=0.05$. The process was simulated on the interval $[0,100]$. }
\label{fig2}  
\end{center}
\end{figure}

In this section we present the model, which is constructed as the superposition of two independent mechanisms, namely the FBM and other process that replaces the time. Here we present two processes that are considered as the "operational time" of FBM, namely the gamma process and the first-exit time of gamma process. The last one we call the inverse gamma process. We should mention here that in the literature  there is considered  an inverse gamma L\'evy process, see e.g. \cite{heyde-leonenko} which is different from the definition of inverse gamma process used in this paper. \\
FBM is one of the most important model considered as the fractional dynamical system. It is well known that this process has self-similarity property and its increments exhibit behavior adequate to long-range dependent systems. Moreover, very often it is considered as a main model appropriate for description of so called anomalous diffusion phenomena. It is worth mentioning that anomalous diffusion property of given process can be recognized by MSD. For a sample $\{X_{i},\; i=1,2,\ldots,n\}$  with stationary increments MSD is defined as follows \cite{burnwer,sierra}
\begin{eqnarray}\label{eq1}
M_n(\tau)=\frac{1}{n-\tau}\sum_{k=1}^{n-\tau}\left(X_{k+\tau}-X_{k}\right)^2.
\end{eqnarray}
The MSD for sample  being a realization of stationary increments process  has the property
\begin{eqnarray}\label{power}
M_n(\tau)\stackrel{\mathcal{L}} = \tau^{2d+1},
\end{eqnarray}
where $\stackrel{\mathcal{L}}=$ means equivalence in law. We classify the process with $d=0$ as exhibiting linear dynamics. If $d<0$, the process may be described as featured by sub-linear dynamics (sub-diffusive process), and consequently $d>0$ points at the super-linear dynamics (super-diffusive) of the stochastic process \cite{sierra}. It is worth to mention, if the sample comes from the H-self similar L\'evy stable process (i.e. process with stationary increments with $\alpha-$stable distribution and H-self similar property), then the $d$ parameter in (\ref{power}) is equal to $d = H-\frac{1}{\alpha}$, \cite{burnwer}. The anomalous diffusion phenomena can also be expressed in language of second moment of given process. More precisely, if $\E(X(t))$ is non-linear function of time, then we call the $X(t)$ process as anomalous diffusive.\\
The FBM was introduced in 1940 by Kolmogorov \cite{NessMandelbrot,b121} and very often is treated as an extension of the classical Brownian motion. Most of the properties of FBM are characterized by the self-similarity exponent $H$, called Hurst exponent.
For any $0<H<1$ the FBM with index $H$ is the mean-zero Gaussian process $B_{H}(t)$ with the following representation \cite{NessMandelbrot,marek}
\begin{eqnarray}\label{FBM}
B_H(t)=\int_{-\infty}^{\infty}\left( (t-u)_{+}^{H-1/2}-(-u)_{+}^{H-1/2}\right)dB(u),\quad t\geq 0,
\end{eqnarray}
where $B(t)$ is the Brownian motion and $(x)_{+}=\max(x,0)$.
It is woth mentioning, the process exhibits subdiffusive dynamics for $H<1/2$ and superdiffusive one for $H>1/2$.
For each $t$, $\mathbb{E}B_H(t) =0$, $\mathbb{E}B^2_H(t) = t^{2H}$ and its probability density function (PDF) is given by
\begin{eqnarray}\label{pFBM}
f_{B_H(t)}(x)=\frac{1}{\sqrt{2\pi} t^H}e^{-\frac{1}{2}x^2 t^{-2H}},~~x\in R.
\end{eqnarray} 
As it was mentioned, the FBM has $H$-self similarity property, which means for all $c>0$ the following holds
\begin{eqnarray}\label{Hself}\{B_{H}(ct),\; t~\geq 0\}\stackrel{\mathcal{L}}=\{c^{H}B_{H}(t),\; t~\geq 0\}.\end{eqnarray}
 One should remember that the increments of FBM, so called fractional Gaussian noise defined as $b_H(n)=B_H(n+1)-B_H(n)$ for $n=0,1,\ldots$, is time-correlated stationary process with covariance function
$$\mbox{Cov}(b_H(0), b_H(n))= \mathbb{E}\left( b_H(0)b_H(n)\right)=\frac{{1}}{2}\left((n+1)^{2H}+(n-1)^{2H}-2n^{2H}\right).$$
Moreover, the Laplace transform of $B_H(t)$ is given by
\begin{eqnarray}
\mathbb{E}\left (e^{-uB_H(t)}\right)=e^{-\frac{1}{2}u^2 t^{2H}}, ~~t \geq 0.
\end{eqnarray}
As the first time-changed process we consider the FBM subordinated by gamma process. This process was considered in \cite{meer} and it is known in the literature as the FLM \cite{codi}. FLM is defined as follows
\begin{eqnarray}\label{flm}
Y_1(t)=B_{H}(U(t)),
\end{eqnarray}
where $B_{H}(t)$ is the FBM with Hurst exponent $H$ and $U(t)$ is the gamma process. We should mention, the gamma process $U(t)$ has stationary independent increments with gamma distribution. More precisely, the increments $U(t+s)-U(t)$ have the following PDF \cite{meer}
\begin{eqnarray}\label{den}
f(x)=\frac{1}{\beta^{\alpha}\Gamma(\alpha)}x^{\alpha-1}e^{-x/\beta},
\end{eqnarray} 
where $\alpha=s/\nu$, $\beta=1$ and $\nu$ is a positive parameter. The main properties of the process were considered in \cite{meer} however in the next section we repeat them in comparison with the properties of the second time-changed process defined as the FBM time-changed by inverse gamma process. In Fig. \ref{fig1} we present the exemplary trajectories of the process $Y_1(t)$ for different values of $H$ and $\nu$ parameters. The details of the simulation procedure are provided in section \ref{sim}. 
\\
The second considered process is the FBM time-changed by inverse gamma process. It is defined as follows
\begin{eqnarray}\label{general}
Y_2(t)=B_{H}(V(t)),
\end{eqnarray}
where $V(t)$ is defined as the inverse process of $U(t)$
\begin{eqnarray}\label{inverse}
V(t)=\inf\{\tau: U(\tau)>t\}.
\end{eqnarray}
In Fig. \ref{fig2} we present the exemplary trajectories of $Y_2(t)$ process for different values of $H$  and $\nu$ parameters. Here we observe constant time periods characteristic for the processes delayed by inverse subordinators. The simulation procedure of $Y_2(t)$ process is presented in section \ref{sim}.
In the literature we can find many interesting  examples of  construction of general subordinated processes similar as in (\ref{general}). Some of the examples are mentioned in section \ref{intro}. Interestingly, the subordinated processes  obtained as in (\ref{general}) exhibit behavior adequate to anomalous diffusive systems and their PDF can be described by the fractional Fokker-Planck equations  \cite{gaj2,klafter}. \\
\section{Main properties of time-changed FBM by gamma and inverse gamma process}\label{theory}
In this section first we discuss the main properties of gamma and inverse gamma processes. Then we explore main properties of $Y_1(t)$ and $Y_2(t)$. Note that the process $Y_1(t)$ is already studied in \cite{meer} and hence we only mention their results for completeness and comparative purpose.
\subsection{Gamma process}
The density function $f(x,t)$ for gamma process $U(t)$ is given by
\begin{equation*}
f(x,t) = \frac{1}{\Gamma(t/\nu)} x^{\frac{t}{\nu}-1}e^{-y},\;y>0,
\end{equation*}
with L\'evy density $\pi_U(x) = e^{-x}/(\nu x)$ (see e.g. \cite{applebaum}, p. 55), which implies
$\int_{0}^{\infty}\pi_U(x) dx=\infty$. By Theorem 21.3 of \cite{sato}, the sample paths of $U(t)$ are strictly increasing with jumps. Further, let $\tilde{f}(u,t) = \mathcal{L}_x(f(x,t)) = \int_{0}^{\infty}e^{-ux}f(x,t)dx$ be the Laplace transform (LT) of $f(x,t)$ with respect to variable $x$, then
\begin{equation}
\tilde{f}(u,t) = (1+u)^{-t/\nu} = e^{-\frac{t}{\nu}\log(1+u)}.
\end{equation}
Thus the Laplace exponent for gamma process $U(t)$ is given by $\Psi_U(u) = \frac{1}{\nu}\log(1+u).$ \\

\begin{prop}\label{moments-gamma}
For $q>0$, the $q$-th order moment of $U(t)$ satisfies $\mathbb{E}(U^q(t)) \sim (t/\nu)^q$ as $t\rightarrow\infty.$
\end{prop}
\noindent{\bf Proof.} Note that as $t\rightarrow\infty,$ it follows
\begin{align*}
\mathbb{E}(U^q(t)) = \frac{\Gamma(q+\frac{t}{\nu})}{\Gamma(\frac{t}{\nu})} \sim \left(\frac{t}{\nu}\right)^q,
\end{align*}
using $\Gamma(n+\alpha)\sim n^{\alpha}\Gamma(n)$ as $n\rightarrow\infty.$
\hfill{\qedsymbol}\\
Since $U(t)$ is a L\'evy process it is straightforward to find the covariance function, namely it is given by Cov$(U(s), U(t)) = \frac{t}{\nu} + \frac{t^2-s^2}{\nu^2},$ where $s<t.$ Further, for fixed $s$ and $t\rightarrow\infty$, it follows Cov$(U(s), U(t))\sim t^2/\nu^2$. It is worth to mention, the right tail of the $U(t)$ process is given by:
\begin{eqnarray}
\mathbb{P}(U(t)>x)=\frac{\Gamma(t/\nu,x)}{\Gamma(t/\nu)}, 
\end{eqnarray}
where $\Gamma(a,z)=\int_z^{\infty}x^{a-1}e^{-x}dx$ is the upper incomplete gamma function.
\subsection{Inverse gamma process}

Let $h(x,t)$ be the density function of $V(t)$ process. 
\noindent It is worth to mention, for a strictly increasing subordinator $Z(t)$ with density function $p(x,t)$ and Laplace exponent $\Psi_Z$, the density function $q(x,t)$ of the first-exit time process has the LT, \cite{meerschaert-scheffler}
\begin{equation}\label{LT-of-inverse-subordinator}
\mathcal{L}_t(q(x,t)) = \frac{1}{s}\Psi_Z(s) e^{-x\Psi_Z(s)}.
\end{equation}
Since $U(t)$ is strictly increasing subordinator with Laplace exponent $\Psi_{U}(u) = \frac{1}{\nu}\log(1+u)$, we obtain from \eqref{LT-of-inverse-subordinator} the LT of $h(x,t)$ with respect to the time variable $t$ is given by
\begin{equation}\label{ch3-prop2.1}
\mathcal{L}_t(h(x,t)) = \frac{1}{\nu}\frac{\log(1+s)}{s(1+s)^{x/\nu}}= Q(x, s)~\rm{(say)}.
\end{equation}
We first obtain an integral representation for $h(x,t)$.
\begin{prop}\label{density-inverse-gamma}
The density function $h(x,t)$ of $V(t)$ is given by 
\begin{equation}
h(x,t) = \frac{e^{-t}}{\nu\pi}\int_{0}^{\infty} \frac{y^{-x/\nu}e^{-yt}}{1+y} \left(\pi \cos\left(\frac{\pi x}{\nu}\right) - \log (y) \sin\left(\frac{\pi x}{\nu}\right)\right) dy,\; x/\nu \not\in \mathbb{N}.
\end{equation}
\end{prop}
\noindent{\bf Proof.}
The density function of $h(x,t)$ can be obtained by using the Laplace inversion formula, \cite{schiff}
\begin{equation}\label{complex-inversion}
h(x,t) = \frac{1}{2\pi i} \int_{x_0-i\infty}^{x_0+i\infty}e^{st}Q(x,s)ds.
\end{equation}

\begin{figure}[ht]
\centering{\includegraphics[scale=0.45]{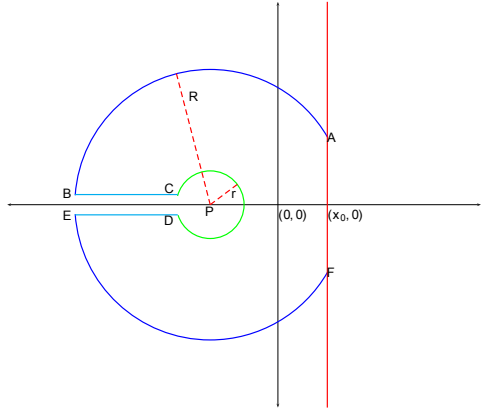}}
\caption{Contour ABCDEFA}
\label{contourplot}
\end{figure}
\begin{figure}[h!]
\begin{center}
\includegraphics[width=0.6\textwidth]{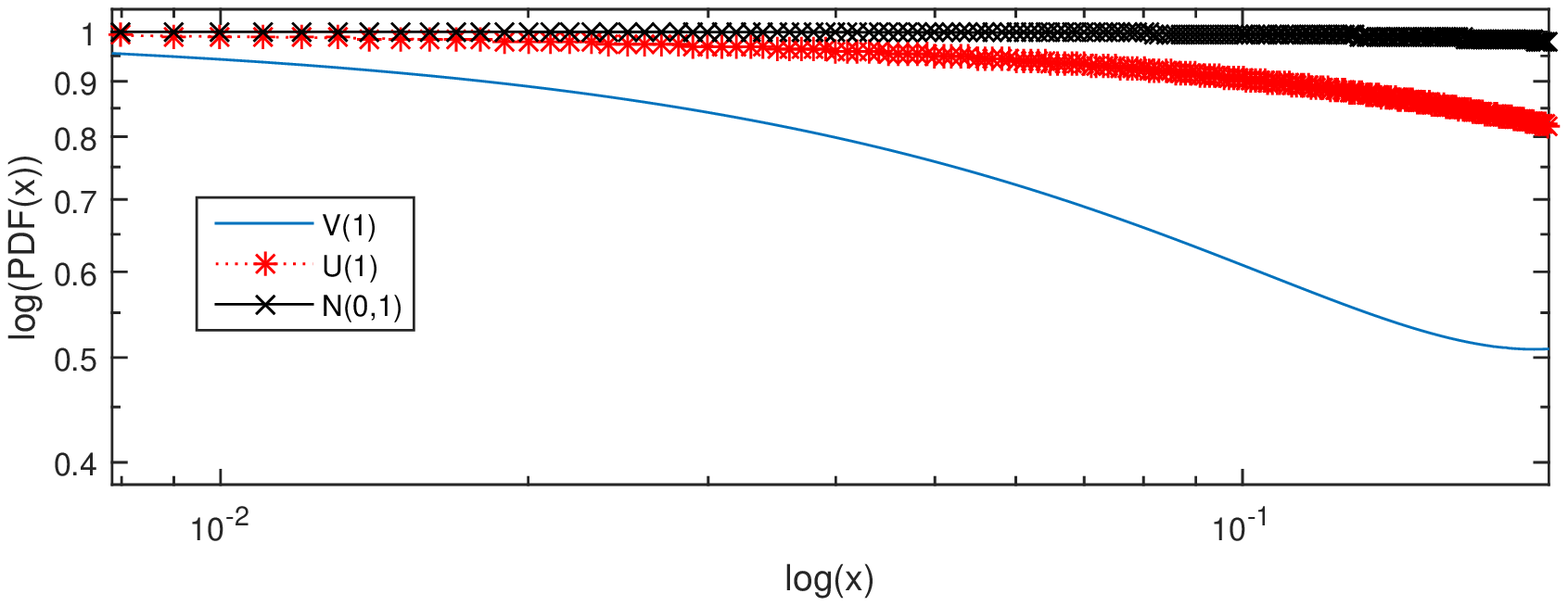}
\caption{The comparison of the PDFs of $U(1)$, $V(1)$ and Gaussian random variables (in log-log scales). }
\label{fig8}  
\end{center}
\end{figure}
\begin{figure}[h!]
\begin{center}
\includegraphics[width=0.6\textwidth]{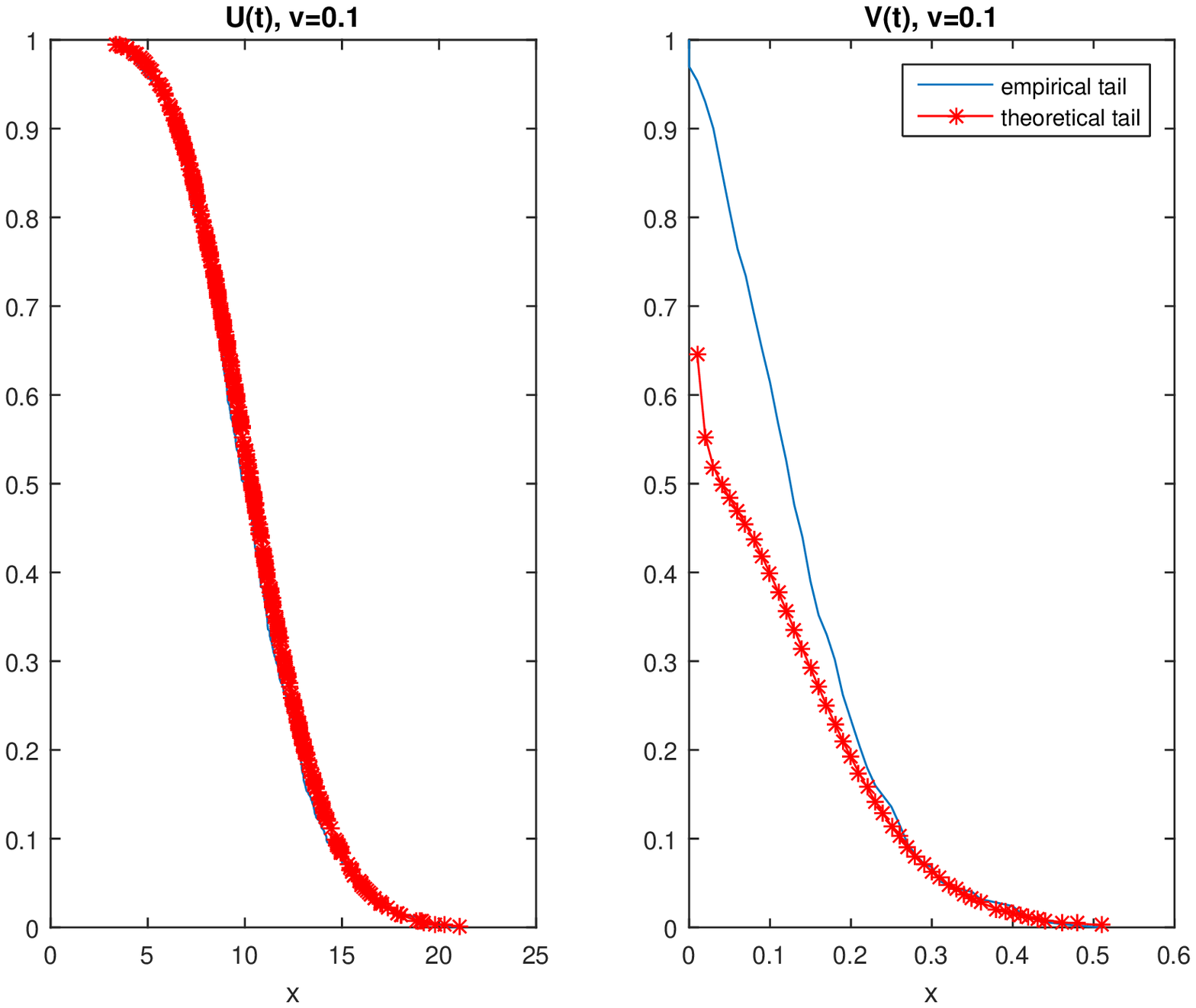}
\caption{Comparison of the tail behavior for gamma process $U(t)$ (left panel) and its inverse $V(t)$ (right panel) for $t=1$. The blue thin line corresponds to the empirical tail calculated on the basis of $500$ trajectories of the corresponding processes while the red star line corresponds to the theoretical tails of appropriate processes.}
\label{fig5}  
\end{center}
\end{figure}

\noindent For calculating integral in \eqref{complex-inversion}, we consider a closed key-hole contour $\mathcal{C}: ABCDEFA$ (see Fig. \ref{contourplot}) with a branch point at P$=(-1, 0)$.
Here AB and EF are arcs of a circle of radius $R$ with center at $P$, BC and DE are line segments parallel to $x$-axis as shown in  Fig. \ref{contourplot}, CD is an arc $\mathcal{C}_r$ of a circle of radius $r$ with center at P and FA is the line segment from $x_0-iy$ to $x_0+iy$ with $x_0>0$. Note that for $x/\nu \not\in \mathbb{N}$, LT $F(x,s)$ has a simple pole at $s=0$ and a branch point at $s=-1$. By residue theorem, we have
\begin{equation}\label{contour-1}
 \begin{split}
\frac{1}{2\pi i}\int_{\mathcal{C}}e^{st}Q(x,s) ds &= \sum  \mbox{Residue} \left(e^{st} Q(x,s)\right) =0,
\end{split}
\end{equation} 
since the residue of $Q$ at simple pole $s=0$, is zero. It is easy to verify that integral along the curve AB and EF vanish as $R\rightarrow\infty.$ Further, integral along arc $\mathcal{C}_r$ also vanishes as $r\rightarrow 0$. Hence using \eqref{complex-inversion} and \eqref{contour-1}, we have
\begin{equation}\label{AB-DE}
h(x,t) = - \lim_{r\rightarrow 0, R\rightarrow \infty} \left(\frac{1}{2\pi i}\int_{BC}e^{st}Q(x,s) ds +\frac{1}{2\pi i}\int_{DE}e^{st}Q(x,s) ds\right).
\end{equation}
\noindent Along BC put $s = ye^{i\pi}$, so that
\begin{align}
\int_{BC} e^{st} Q(x,s)ds &= \frac{1}{\nu}\int_{-R-1}^{-r-1}e^{st}\frac{\log(1+s)}{s(1+s)^{x/\nu}}ds\nonumber\\
& = \frac{e^{-t}}{\nu}\int_{R}^{r}\frac{e^{-yt}(\log(y) +i\pi)} {(y+1)y^{x/\nu}} e^{-i\pi x/\nu} dy.
\end{align}
\noindent Further, along DE put $s = ye^{-i\pi}$, which implies
\begin{align}
\int_{DE} e^{st} Q(x,s) ds&= \frac{1}{\nu}\int_{-r-1}^{-R-1}e^{st}\frac{\log(1+s)}{s(1+s)^{x/\nu}}ds\nonumber\\
& = \frac{e^{-t}}{\nu}\int_{r}^{R}\frac{e^{-yt}(\log(y) -i\pi)} {(y+1)y^{x/\nu}} e^{i\pi x/\nu} dy.
\end{align}
\noindent Thus,
\begin{align}\label{final-exp}
\int_{BC} e^{st} Q(x,s)ds + \int_{DE} e^{st} Q(x,s) ds = (2i)\frac{e^{-t}}{\nu}\int_{r}^{R} \left(\log (y) \sin\left(\frac{\pi x}{\nu}\right)-\pi \cos\left(\frac{\pi x}{\nu}\right) \right) dy.
\end{align}
The result follows using \eqref{AB-DE} and \eqref{final-exp}.
\hfill{\qedsymbol}\\
In order to illustrate the above result, in Fig. \ref{fig8} we present the PDF (for large $x$) of the processes $V(t)$ and $U(t)$ for $t=\nu=1$ (in log-log scale). Moreover, as a comparison we show also the PDF of standard normal distribution.\\
Next we obtain the tail probability behavior for $V(t)$. Let $\gamma(a,z) = \int_{0}^{z} y^{a-1}e^{-y}dy$ be lower incomplete gamma function. Then from \cite{abramowitz} it follows
\begin{equation}\label{incomplete-gamma}
\frac{\gamma(a,z)}{\Gamma(a)} \sim (2\pi a)^{-1/2} e^{a-z} \left(\frac{z}{a}\right)^a\;\; \mbox{for large $a$ and fixed $z$}.
\end{equation}
\begin{prop}
The tail of inverse gamma process $V(t)$ satisfies
\begin{equation}
\mathbb{P}(V(t)> x) \sim \left(\frac{2\pi x}{\nu}\right)^{-1/2} e^{x/\nu-t}\left(\frac{\nu t}{x}\right)^{x/\nu},\;\; \mbox{as}\; x\rightarrow \infty.
\end{equation}
\end{prop}
\noindent{\bf Proof.}
Note that
\begin{align*}
\mathbb{P}(V(t) > x) = \mathbb{P}(U(x) \leq t)& = \int_{0}^{t} \frac{1}{\Gamma(x/\nu)}y^{x/\nu-1}e^{-y}dy\\
& = \frac{1}{\Gamma(x/\nu)}\gamma(x/\nu, t).
\end{align*}
The result follows using \eqref{incomplete-gamma}.
\hfill{\qedsymbol}\\
In Fig. \ref{fig5} we present the comparison of the  empirical right tails of $U(t)$ (left panel) and $V(t)$ (right panel) for $t=1$ together with their theoretical tails. To the simulations we take $500$ trajectories of the corresponding processes. The $\nu$ parameter in both cases is equal to $1$.
\begin{prop} \label{moments-ig}
For $q>0$, the $q$-th order moment of $V(t)$ satisfies $\mathbb{E}(V^q(t)) \sim (\nu t)^q$ as $t\rightarrow\infty.$
\end{prop}
\noindent{\bf Proof.} Let $M_q(t) = \mathbb{E}(V^q(t))$. Let $\tilde{M}_{q}(s)$ be the LT of $M_q(t)$. From \cite{kumar}, we have 
\begin{align*}
\tilde{M}_{q}(s) &= \frac{\Gamma(1+q)}{s(\Psi_U(s))^q} = \frac{\nu^q \Gamma(1+q)}{s(\log(1+s)^q)}\\
&\sim \frac{\nu^q\Gamma(1+q)}{s^{q+1}}\;\mbox{as}\; s\rightarrow 0.
\end{align*}
The result follows now by Tauberian theorem, see e.g. \cite{bertoin}, p. 10.
\hfill{\qedsymbol}\\
We provide here exact form of first and second moments of $V(t)$ i.e. $M_1(t)$ and $M_2(t)$. Note that $\tilde{M}_1(s) = \nu/(s \log(1+s)).$
Using a similar contour and the same argument as in Proposition \ref{density-inverse-gamma} for complex inversion, we obtain 
\begin{equation}
M_1(t) = \mathbb{E}V(t) = \nu\left(t+\frac{1}{2}\right) - \nu e^{-t} \int_{0}^{\infty} \frac{e^{-yt}}{(1+y) [(\log(y))^2+\pi^2]}dy.
\end{equation}
Further, $\tilde{M}_2(s) = \nu^2/(s [\log(1+s)]^2)$, which yields after Laplace inversion
\begin{equation}
M_2(t) = \mathbb{E}V^2(t) = \nu^2\left(\frac{1}{12}+t+\frac{t^2}{2}\right) - 2\nu^2e^{-t}\int_{0}^{\infty} \frac{e^{-yt}}{(1+y) } \frac{\log(y)}{\left((\log(y))^2+\pi^2\right)^2}dy.
\end{equation}
\noindent Using \cite{mijena} we can obtain that the covariance structure for $V(t)$ is given by
\begin{equation*}
\mbox{Cov}(V(s), V(t)) = \frac{1}{2}M_2(s) + \int_{0}^{s}M_1(t-y)dM_1(y) -M_1(t)M_1(s).
\end{equation*}
\subsection{FBM time-changed by gamma process}
Gamma process is a driftless, strictly increasing and pure jump L\'evy process and hence the sample paths of the subordinated process $Y_1(t) = B_H(U(t))$ also have jumps. Further, $Y_1(t)$ is a non-Gaussian, non-Markovian process with stationary but dependent increments. Note that $B_H(1)$ is normal random variable with mean 0 and variance 1, and hence $\mathbb{E}|B_H(1)|^q = \sqrt{\frac{2^q}{\pi}}\Gamma\left(\frac{1+q}{2}\right) = c_q$ (say). By using this fact and self-similarity of FBM with Proposition \ref{moments-gamma}, the asymptotic behaviors of moments of $Y_1(t)$ satisfy, see \cite{meer}
\begin{equation}\label{asym-mom-flm}
\mathbb{E}|Y_1(t)|^q = c_q \frac{\Gamma(Hq+t/\nu)}{\Gamma(t/\nu)} \sim c_q (t/\nu)^{qH},\;\mbox{as}\;t\rightarrow \infty. 
\end{equation}
Because the second moment of $Y_1(t)$ process is non-linear for large $t$ we conclude the FBM time-changed by gamma process can be classified as anomalous diffusive system.\\
\noindent The density function of $Y_1(t)$ is given by (see \cite{meer}, Proposition 2.5)
\begin{equation*}
f_{Y_1(t)}(x) = \frac{1}{\sqrt{2\pi}\Gamma(t/\nu)}\int_{0}^{\infty} y^{t/\nu-H-1} e^{-\frac{1}{2}x^2y^{-2H}-y} dy,\;\;x\neq 0.
\end{equation*}
Further, the covariance function for $Y_1(t)$ follows, \cite{meer}
\begin{equation}\label{cov-flm}
\mathbb{E}Y_1(s)Y_1(t) = \frac{1}{2}\left( \frac{\Gamma(2H+t/\nu)}{\Gamma(t/\nu)} + \frac{\Gamma(2H+s/\nu)}{\Gamma(s/\nu)} - \frac{\Gamma(2H+(t-s)/\nu)}{\Gamma((t-s)/\nu)}\right),\; s<t.
\end{equation}
Next we show that the process $Y_1(t)$ has LRD property. It is worth to mention here that the LRD for stationary increment process of $Y_1(t)$ is discussed in \cite{meer}. However, we establish here LRD for process $Y_1(t)$ itself which is non-stationary. A finite variance stationary process $X(t)$ is said to have LRD property, \cite{cont}, if $\sum_{k=0}^{\infty}\gamma_k = \infty$, where
\begin{equation*}
\gamma_k = \mbox{Cov}(X(t), X(t+k)).
\end{equation*}
Further, for a non-stationary process $X(t)$ an equivalent definition is given by
\begin{definition}\label{lrd-def}
Let $s>0$ be fixed and $t>s$. Then process $X(t)$ is said to have LRD property if
\begin{equation}
\rm{Corr}(X(s), X(t) ) \sim c(s) t^{-d},\;\;\rm{as}\;t\rightarrow\infty,
\end{equation} 
where $c(s)$ is a constant depending on $s$ and $d\in(0,1)$. 
\end{definition}
\noindent Let $g(x) = \Gamma(x + 2H)/\Gamma(x)$. By the Taylor expansion, we have \cite{meer}
\begin{equation}\label{asym-behv}
\frac{g(x+h)}{g(x)} = 1 + 2H(h/x) + H(2H-1)(h/x)^2 + \mbox{O}(x^{-3}).
\end{equation}
For fixed $s$ and large $t$, using \eqref{cov-flm} and \eqref{asym-behv}, it follows
\begin{align*}
\mathbb{E}Y_1(s)Y_1(t) &= \frac{g(t/\nu)}{2}\left( 1+ \frac{g(s/\nu)}{g(t/\nu)} - \frac{g((t-s)/\nu)}{g(t/\nu)}\right)\\
& = \frac{g(t/\nu)}{2}\left[ 1+ \frac{g(s/\nu)}{g(t/\nu)} - \left(1 - 2H\left(\frac{s}{t}\right) + H(2H-1)\left(\frac{s^2}{t^2}\right) + \mbox{O}(t^{-3}) \right)\right]\\ 
&\sim \frac{2H s}{\nu^{2H}} t^{2H-1},\;\;\mbox{as}\;t\rightarrow \infty. 
\end{align*}
\noindent Thus we have following result for asymptotic behavior of covariance structure of FBM time-changed by gamma process.
\begin{prop}
For fixed $s$ and as $t\rightarrow \infty$, 
\begin{equation*}
\mathbb{E}Y_1(s) Y_1(t) \sim \frac{2H s}{\nu^{2H}} t^{2H-1}.
\end{equation*}
\end{prop}

\noindent Further, using \eqref{asym-mom-flm}, it follows
\begin{align}\label{corr-tcfbm1}
\mbox{Corr}(Y_1(s), Y_1(t))\sim \frac{2H s}{\nu^h \sqrt{\mathbb{E}Y^2_1(s)}}t^{H-1},\;\;\mbox{as}\;t\rightarrow \infty
\end{align}
\noindent Using Definition \ref{lrd-def} and Eq. \eqref{corr-tcfbm1} we have following result.
\begin{prop}
Time-changed FBM by  gamma process has LRD property for every $H$.
\end{prop}

\subsection{FBM time-changed by inverse gamma process}
\begin{figure}[h!]
\begin{center}
\includegraphics[width=0.6\textwidth]{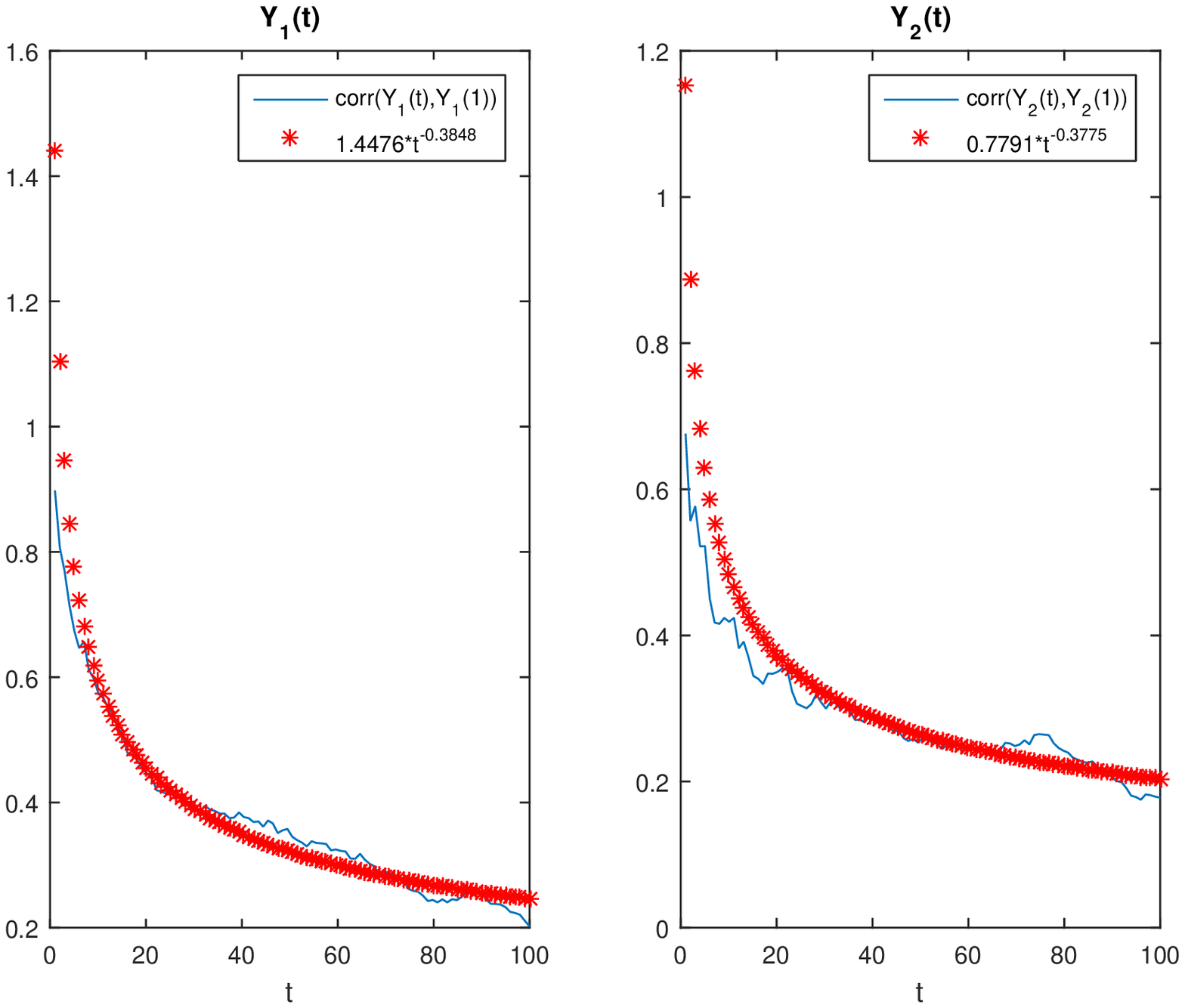}
\caption{The empirical correlation $\mbox{Corr}(Y_1(s), Y_1(t))$ as a function of $t$ for $s=1$ (left panel) and   $\mbox{Corr}(Y_2(s), Y_2(t))$ (right panel) for  $H=0.7$ and $\nu=1$. The red star lines correspond to  the power functions fitted by using the least squares method.}
\label{fig9}  
\end{center}
\end{figure}

In this subsection, we discuss the main properties of FBM time-changed by inverse gamma process. Note that the sample paths of $U(t)$ are strictly increasing with jumps, and hence the sample paths of $V(t)$ are almost surely continuous and are constant over the intervals where $U(t)$ have jumps. Hence the sample paths of $Y_2(t)$ will also be continuous. Moreover, $V(t)$ neither have independent nor stationary increments and hence $Y_2(t)$ is non-Markovian, non-Gaussian process with neither independent nor stationary increments. Using Proposition \ref{moments-ig} and self-similarity of FBM, it follows that
\begin{equation*}
\mathbb{E}|Y_2(t)|^q  \sim c_q (\nu t)^{qH},\;\;\mbox{as}\;t\rightarrow\infty.
\end{equation*}
That means the asymptotic behaviors of $q$-th moments of the $Y_1(t)$ and $Y_2(t)$ are similar. Moreover, $Y_2(t)/(\nu t)^{H} \stackrel{d}= V(t)^HB_H(1)/(\nu t)^{H} \stackrel{d}\rightarrow B_H(1) $ as $t\rightarrow\infty,$ since $V(t)/(\nu t) \stackrel{d}\rightarrow 1$ as $t\rightarrow\infty$, which reflects the fact
that the increments of the $Y_2(t)$ become Gaussian with an increasing lag. Further, the density function of $Y_2(t)$ is given by
\begin{align*}
f_{Y_2(t)}(x) & = \int_{0}^{\infty} f_{B_H(y)}(x)h(y,t) dy\\
&= \int_{0}^{\infty} \frac{1}{\sqrt{2\pi}y^H} e^{-\frac{1}{2} x^2 y^{-2H}}h(y,t) dy,\;\; x\neq 0.
\end{align*}
The covariance structure for time-changed FBM is discussed in \cite{mijena}. However they haven't explicitly discussed the covariance structure for time-changed FBM by inverse gamma process. Here we provide an explicit asymptotic behavior for covariance structure of $Y_2(t)$. The covariance function for $Y_2(t)$ for $s<t$ is given by (see Theorem 3.1. \cite{mijena})
\begin{equation*}
\mathbb{E}Y_2(s)Y_2(t) = M_{2H}(s) + 2H \int_{0}^{s}M_{2H-1}(t-y)dM_1(y).
\end{equation*}
For fixed $s$ and large $t$, we have
\begin{align*}
\int_{0}^{s}M_{2H-1}(t-y)dM_1(y) &\sim \nu^{2H-1}\int_{0}^{s}(t-y)^{2H-1}M_1'(y)dy\\
&\sim \nu^{2H-1}(t-s)^{2H-1}M_1(s) \sim \nu^{2H-1}t^{2H-1}M_1(s).
\end{align*}
Thus we have following result for asymptotic behavior of covariance structure of FBM time-changed by inverse gamma process.
\begin{prop}
For fixed $s$ and as $t\rightarrow \infty$, 
\begin{equation*}
\mathbb{E}Y_2(s)Y_2(t) \sim M_{2H}(s)+ (2H)\nu^{2H-1}t^{2H-1}M_1(s).
\end{equation*}
\end{prop}
\noindent Note that Var$(Y_2(t)) = \mathbb{E}Y^2_2(t) = \mathbb{E}V(t)^{2H} \sim (\nu t)^{2H}$, as $t\rightarrow\infty.$ Hence, for fixed $s$ 
\begin{equation}\label{corr-tcfbm}
\mbox{Corr}(Y_2(s), Y_2(t))\sim \frac{M_{2H}(s)}{\sqrt{\mathbb{E}Y^2_2(s)}}(\nu t)^{-H}+\frac{2H M_1(s)\nu^{H-1}}{ \sqrt{\mathbb{E}Y^2_2(s)}}t^{-(1-H)},\;\; \mbox{as}\; t\rightarrow\infty.
\end{equation}
\noindent Using Definition \ref{lrd-def} and Eq. \eqref{corr-tcfbm} we have following result.
\begin{prop}
Time-changed FBM by inverse gamma process has LRD property for every $H$.
\end{prop}
\noindent Because the second moment of $Y_2(t)$ process is non-linear function for large $t$ then we conclude the process is anomalous diffusive.\\
In Fig. \ref{fig9} we present the empirical correlation $\mbox{Corr}(Y_1(s), Y_1(t) )$ as a function of $t$ for $s=1$ (left panel) while in the right panel we present  $\mbox{Corr}(Y_2(s), Y_2(t))$. To the analysis we take $500$ trajectories of both considered processes with $H=0.7$ and $\nu=1$. On both panels we also present the power functions fitted by using the least squares method. 

\section{Simulation procedures}\label{sim}
\subsection{FBM time-changed by gamma process}
The simulation procedure of the process $Y_1(t)$ is given in \cite{meer}, however in this section we mention it. The main idea is to simulate independent trajectories of the subordinator $U(t)$ and the FBM $B_{H}(t)$. By taking their superposition we obtain the trajectory of the process $Y_1(t)$ defined in (\ref{flm}).
First we simulate the trajectory of $U(t)$, that is a process of independent stationary increments having gamma distribution.  We divide the interval $[0,T]$   into  sub-intervals of length $\delta$ where the increments $U(t+\delta)-U(t)$, $t=0,\delta ,2\delta ,\ldots,T-\delta $ have gamma distribution with parameters $\delta/\nu$ and $1$. We simulate $[T/\delta]$ independent random variables from this distribution. Finally, the trajectory of $U(t)$ is obtained as the cumulative sum of the increments.\\
Secondly, we simulate the approximate trajectory of the external process - FBM which is a special case of fractional L$\acute{e}$vy-stable motion with index of stability equal to $2$.  The method was introduced in \cite{m60} and is based on the generation of an approximate path of the stationary process $b_H(\delta n )=B_{H}(\delta(n+1) )-B_{H}(\delta n)$, $n=0,1,2,\ldots $. Its main idea is to approximate the stochastic integral
\begin{eqnarray*}
b_H(\delta n) =\int_{-\infty}^{\infty}\left( (\delta n-u)_{+}^{H-1/2}-(\delta n-u-1)_{+}^{H-1/2}\right)dB(u),\quad n=0,1,2,\ldots
\end{eqnarray*}
by the Riemann sum
\begin{equation}\label{Riemann}
Y_{m,M}(\delta n)=\sum_{j=1}^{mM}\left((j/m)_{+}^{H-1/2}-(j/m-1)_{+}^{H-1/2}\right)\tilde{B}_{m}(m\delta n-j),\quad n=0,1,2,...
\end{equation}
where $\tilde{B}_{m}(j)=B((j+1)/m)-B(j/m)$. The parameters $m$ and $M$ control the mesh size and the kernel function cut-off, respectively. For large values of these parameters the variables $Y_{m,M}(\delta n)$ approximate well the variables $b_H(\delta n) $ \cite{m60}. But the direct implementation of the sum (\ref{Riemann}) yields a slow algorithm and can not be efficiently applied to generate long approximations in real time. Therefore the method presented in \cite{m60} uses the fast Fourier transform algorithm and the technique of discrete Fourier transforms for some circular sequences. The details can be found in \cite{m60}. The trajectory of FBM $B_{H}(t)$ can be obtained as cumulative sums of variables $b_H(\delta n)$. The similar pocedure applied to general L\'evy-stable motion is presented in \cite{marek}.
Finally, by taking superposition of trajectories of $B_{H}(t)$ and $U(t)$ we obtain  approximated trajectory of process $Y_1(t)$. The exemplary trajectories of the process $Y_1(t)$ obtained by using the described procedure are presented in Fig. \ref{fig1}.
\subsection{FBM time-changed by inverse gamma process}
The idea of simulation of $Y_2(t)$ process is similar as discussed above. It is also based on the simulation of independent trajectories of FBM and the process $V(t)$ which is an inverse to the gamma process $U(t)$. In  order to simulate the approximate trajectory of  the inverse gamma subordinator  first we need to define $V_\delta(t)$ with the step length $\delta$ in the following way
$$
V_\delta(t)=\left(\min\{n\in\mathbb{N}:U(\delta n)>t\}-1\right)\delta,\;\; n=1,2,\ldots,
$$
where $U(\delta n)$ is the value of gamma subordinator $U(t)$ evaluated at $\delta n$, which can be simulated by using the method presented above. Observe that trajectory $V_\delta(t)$ has increments of length $\delta$ at random time instants governed by process $U(\tau)$ and therefore $V_\delta(t)$ is approximation of operational time. Finally, the trajectory of the process $Y_2(t)$ is obtained as the superposition of FBM, for which the simulation procedure is mentioned above, and the process $V(t)$. The exemplary trajectories of the process $Y_2(t)$ obtained by using described procedure are presented in Fig. \ref{fig2}.

\section{Estimation procedures}\label{est}

\subsection{FBM time-changed by gamma process}
\begin{figure}[h!]
\begin{center}
\includegraphics[width=0.6\textwidth]{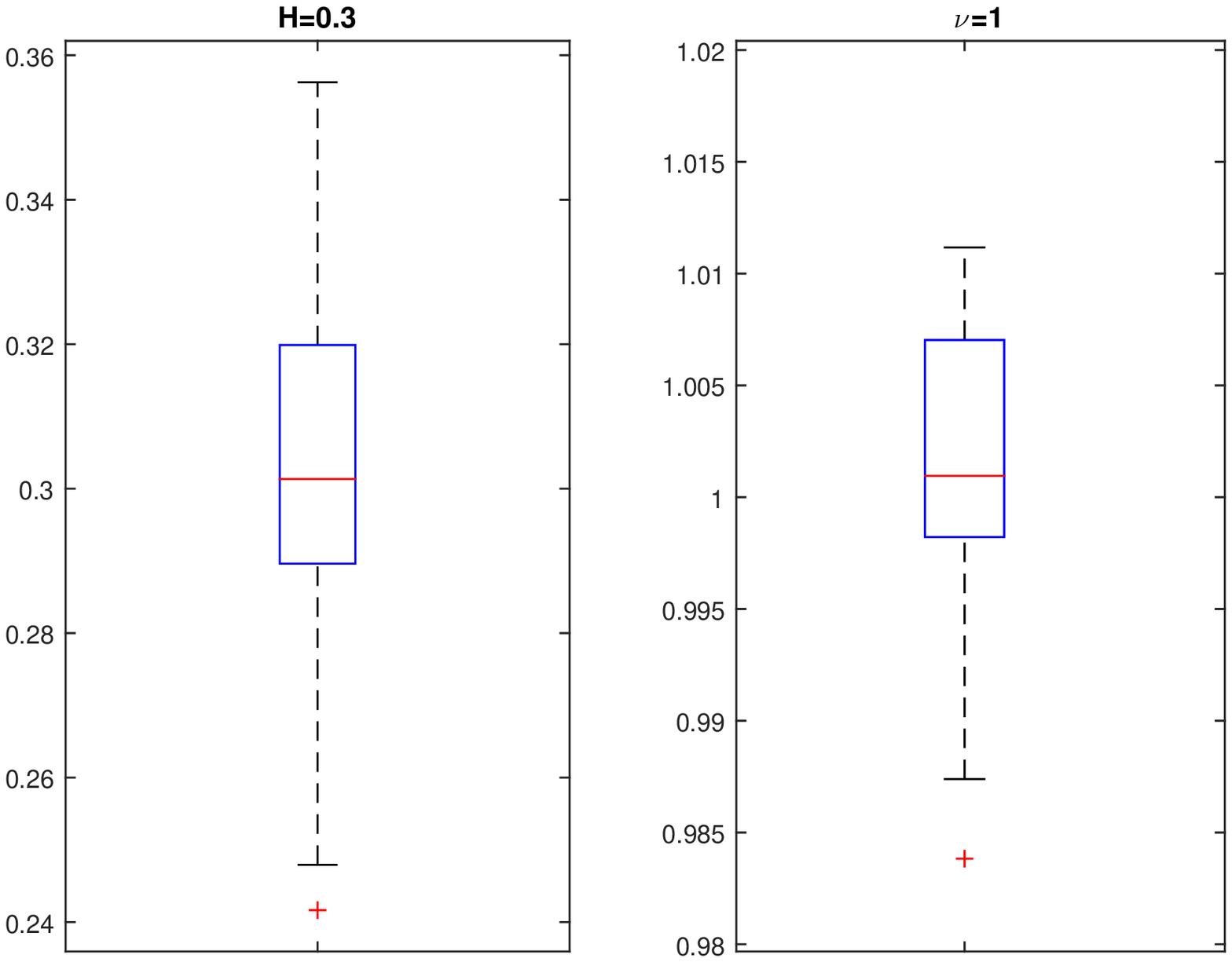}
\caption{The boxplots of estimated $H$ parameter (left panel) and $\nu$ parameter corresponding to $Y_1(t)$ process, i.e. FBM time-changed by gamma process. The theoretical values are $H=0.3$ and $\nu=1$. }
\label{fig3}  
\end{center}
\end{figure}

\begin{figure}[h!]
\begin{center}
\includegraphics[width=0.6\textwidth]{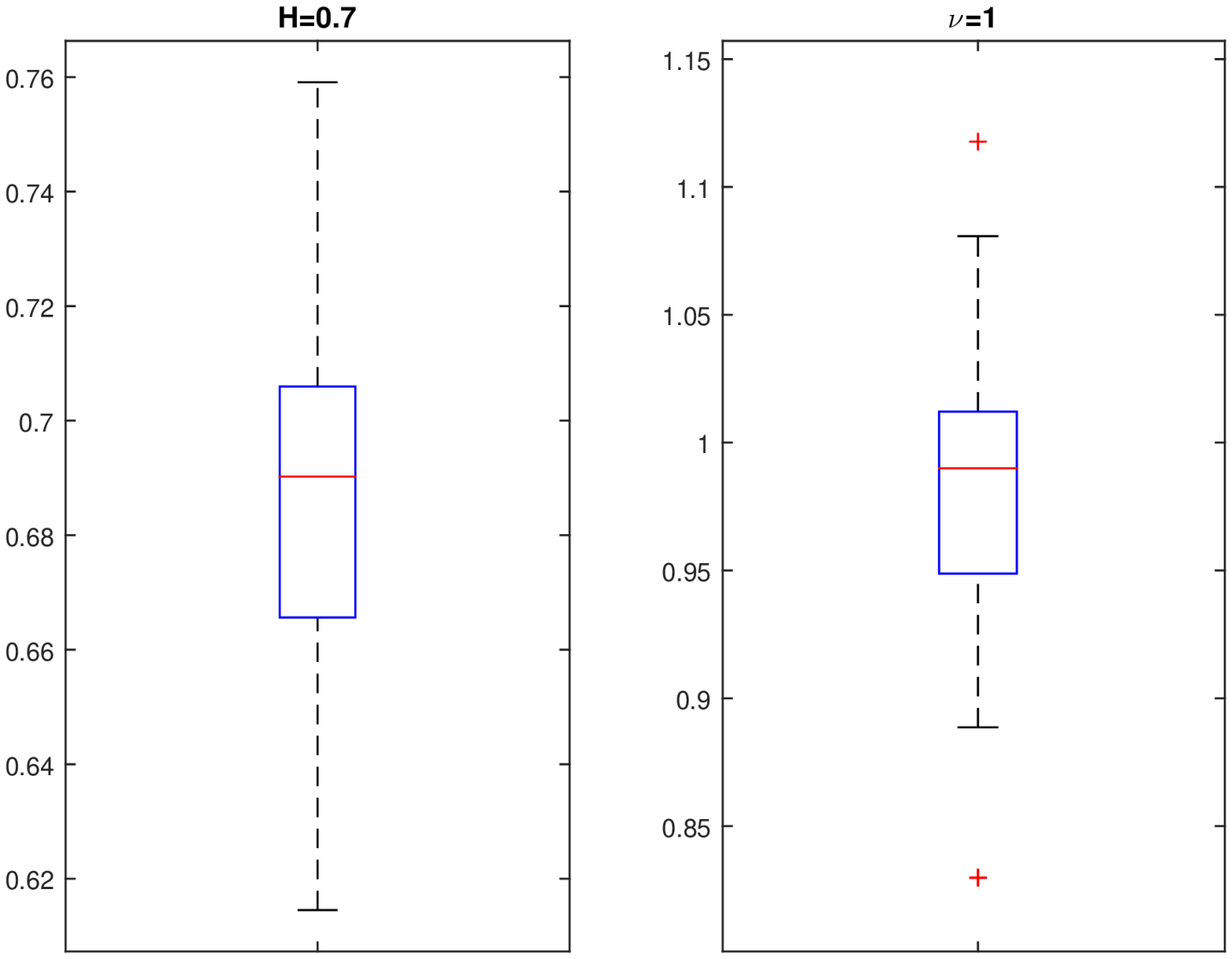}
\caption{The boxplots of estimated $H$ parameter (left panel) and $\nu$ parameter corresponding to $Y_1(t)$ process, i.e. FBM time-changed by  gamma process. The theoretical values are $H=0.7$ and $\nu=1$. }
\label{fig4}  
\end{center}
\end{figure}

The estimation of the Hurst exponent $H$ of FBM time-changed by gamma process is based on the fact that the MSD for process $Y_1(t)$ has similar asymptotic behavior as for the classical FBM, i.e. it satisfies condition (\ref{power}). Using self-similarity and stationarity of increments of FBM, we have
\[M_n(\tau)=\frac{1}{n-\tau}\sum_{k=1}^{n-\tau}\left(Y_1(k+\tau)-Y_{2}(k)\right)^2\stackrel{\mathcal{L}}=\frac{1}{n-\tau}\sum_{k=1}^{n-\tau}U(\tau)^{2H}B_{H}(1)^2\stackrel{\mathcal{L}}= \tau^{2d+1},\]
where $d=H-1/2$.\\
By using this fact for the observed values corresponding to the process $Y_1(t)$ we calculate a sample MSD  for appropriate values of $\tau$ parameter and compare it to the power function $\tau^{2d+1}$. By using the least squares method we estimate the $d$ parameter. Finally we obtain the $H$ parameter which is equal to $d+1/2$. The parameter $\nu$ corresponding to gamma distribution of $U(t)$ process can be estimated on the basis of the behavior of the right tail of the process $Y_1(t)$ proved in \cite{meer}. It should be mentioned, for  estimation of the $v$ parameter we need to have a number of trajectories of the process $Y_1(t)$. We denote them $Y_{11}(t), Y_{12}(t),\ldots,Y_{1K}(t)$. For fixed $t_{0}$ the tail of $Y_1(t_{0})$ satisfies
\begin{equation*}
\mathbb{P}(Y_1(t_0)>x)=a(t_0,H)x^{2(t_0/\nu-1)/(1+2H)}\exp(-b(t_0,H)x^{2/(1+2H)}),~x\rightarrow \infty,
\end{equation*}
where $a,b$ are constants depending on $t_0$ and $H$. Therefore after estimation of $H$ parameter (for one of the trajectory) by comparing the empirical tail  calculated on the basis of $Y_{11}(t_{0}), Y_{12}(t_{0}),...,Y_{1K}(t_{0})$ and theoretical one we can estimate the parameter $\nu$ by using the least squares method, i.e. the estimated parameter minimizes the distance between empirical tail and the theoretical one.  We remind the reader the empirical tail is defined as $1-\hat{F}_n(t)$, where $\hat{F}_n(t)$ is the empirical cumulative distribution function that for vector of observations $x_1,x_2,...,x_n$ has the form
\begin{eqnarray}
\hat{F}_n(t) = \frac{1}{n}\sum_{i=1}^{n}{\textbf{1}_{x_i\leq t}},
\label{empCDF}
\end{eqnarray}
where ${\textbf{1}_{A}}$ is the indicator of the set $A$.\\
In order to check the effectiveness of the described estimation procedure we  simulate $100$ trajectories of length $1000$ of the process $Y_1(t)$ and for each of them we estimate the parameters $H$ and $\nu$. Finally, we create the boxplots of the obtained estimators. The boxplot provides a statistical information about the distribution of the analyzed values \cite{boxplot}. Precisely, it produces a box and whisker plot for each estimated value of $H$ and $v$ parameters. The box has lines at the lower quartile ($Q1)$, median, and upper quartile ($Q3$) values. The whiskers are lines extending from each end of the box to show the extent of the rest of the data. 
Points are drawn as outliers if they are larger than $Q3+1.5(Q3-Q1)$ or smaller than $Q1-1.5(Q3-Q1)$, where $Q1$ and $Q3$ are lower and upper quartiles, respectively \cite{krzysiek}. \\
In Fig. \ref{fig3} we present the boxplot for estimated $H$ parameter (left panel) and $\nu$ parameter (right panel) calculated on the basis of trajectories of $Y_1(t)$ process with theoretical values $H=0.3$ and $\nu=1$. In Fig. \ref{fig4} we show the estimation results for $Y_1(t)$ process with theoretical values $H=0.7$ and $\nu=1$.
The boxplots for $\nu$ parameter are calculated on the basis of estimated values of the parameter for $t=1$ therefore in order to create boxplots we simulate $20$ times  $100$ trajectories of the process $Y_1(t)$ and for observations corresponding to $t=1$ we estimate the appropriate parameter. As we observe in Figs. \ref{fig3} and \ref{fig4} the estimated values of appropriate parameters coincide with the theoretical ones which indicates the proposed estimation procedure can be applied to real data analysis.
\subsection{FBM time-changed by inverse gamma process}
\begin{figure}[h!]
\begin{center}
\includegraphics[width=0.6\textwidth]{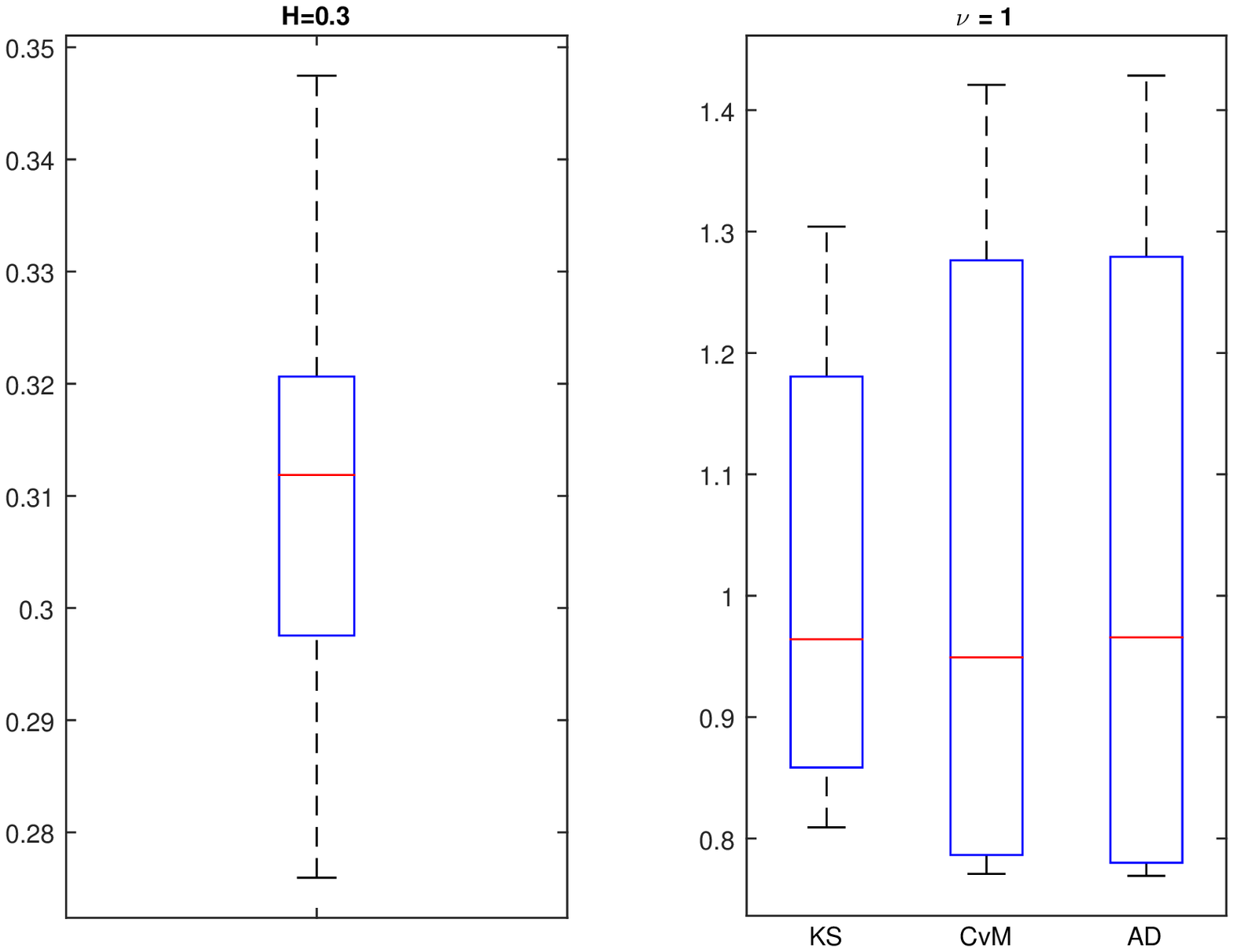}
\caption{The boxplots of estimated $H$ parameter (left panel) and $\nu$ parameter corresponding to $Y_2(t)$ process, i.e. FBM time-changed by inverse gamma process. The theoretical values are $H=0.7$ and $\nu=1$. }
\label{fig6}  
\end{center}
\end{figure}

\begin{figure}[h!]
\begin{center}
\includegraphics[width=0.6\textwidth]{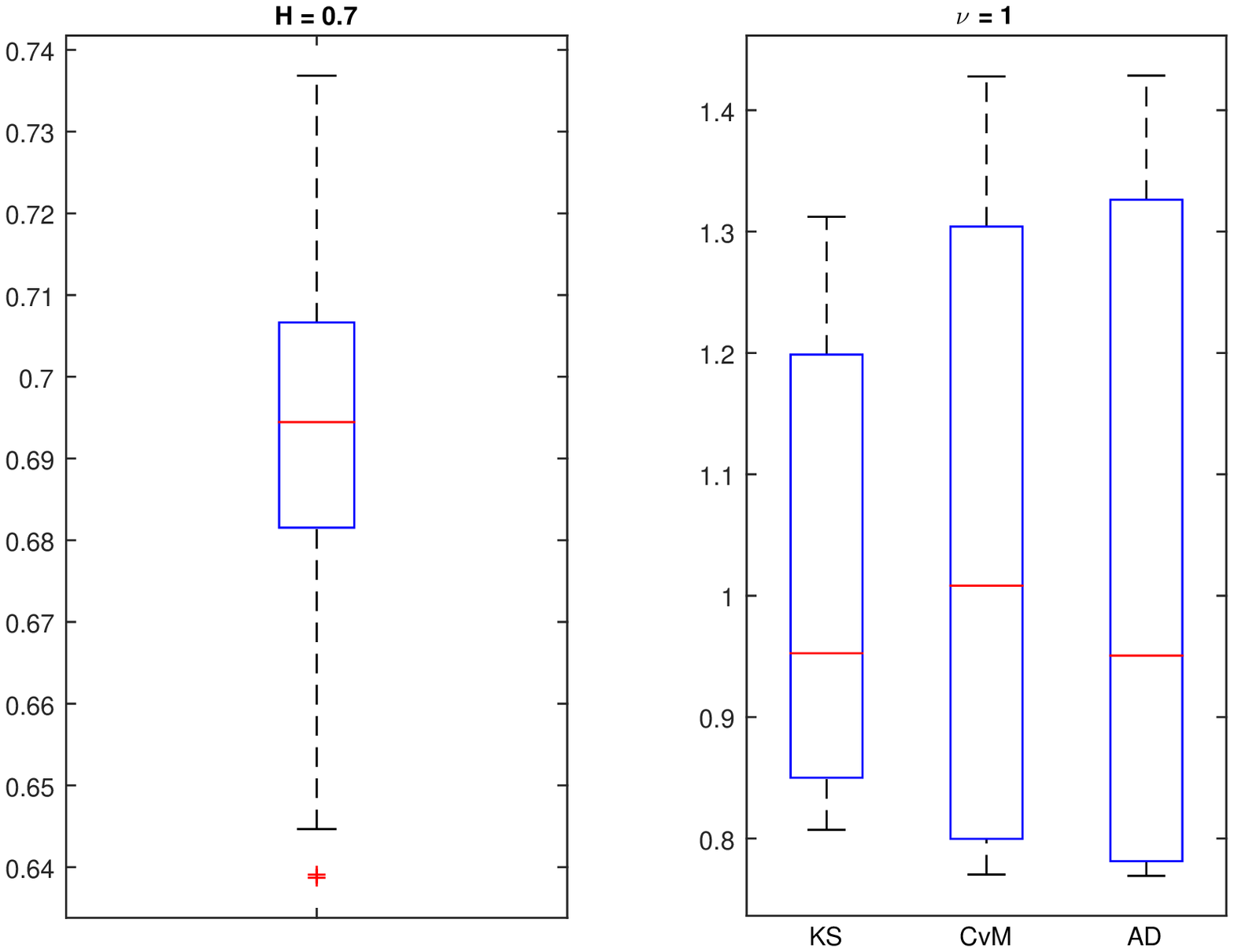}
\caption{The boxplots of estimated $H$ parameter (left panel) and $\nu$ parameter corresponding to $Y_2(t)$ process, i.e. FBM time-changed by inverse gamma process. The theoretical values are $H=0.7$ and $\nu=1$. }
\label{fig7}  
\end{center}
\end{figure}

The estimation procedure of the parameters of FBM driven by inverse gamma subordinator  is divided into two steps. Here we use important property of the process $Y_2(t)$, namely, constant time periods in trajectories which correspond to the jumps of the process $U(t)$. In first step we divide the analyzed time series into two vectors. The first one (vector ${U}$) represents  lengths of constant time periods observed in the data which means the number of consecutive observations that are on the same level. According to the idea of constructing the inverse subordinators, the vector ${U}$ constitutes sample of independent identically distributed (i.i.d.) random variables from the same distribution as the subordinator, i.e. in our case - from gamma distribution. The second vector (vector ${B_{H}}$) arises after removing the constant time periods. This vector constitutes a trajectory of FBM. This scheme is a standard procedure in the analysis of the processes subordinated by inverse subordinators and was used for many applications, see for instance \cite{orzel_ja, janusz_ja}.\\
In the second step we separately analyze the vectors ${U}$ and ${B_{H}}$. On the basis of constant time periods we estimate the parameter $\nu$ from the gamma distribution given by PDF (\ref{den}) for $\alpha=s/\nu$ and $\beta=1$. 
The proposed methodology of fitting the distribution parameter $v$ corresponding to gamma process is based on the minimum distance estimation applied to a gamma distribution. This procedure was proposed in \cite{rafal} for different inverse subordinators however in this paper we sketch their idea. Let $K$ and $L$ denote two functions with a common support on $\mathbb{R}$, the considered distances are
\begin{itemize}
	\item Kolmogorov-Smirnov (KS)
	\begin{eqnarray*}
	KS(K,L) = \sup_{x \in R}{\left|K(x)-L(x)\right|}
	\end{eqnarray*}
	\item Cram\'er-von Mises (CvM)
	\begin{eqnarray*}
	CvM(K,L) = \int_{-\infty}^{\infty}(K(x)-L(x))^2 dL(x)
	\end{eqnarray*}
	\item Anderson-Darling (AD)
	\begin{eqnarray*}
	AD(K,L) = \int_{-\infty}^{\infty} \frac{(K(x)-L(x))^2}{L(x)(1-L(x))} dL(x).
	\end{eqnarray*}
\end{itemize}
In our estimation procedure we consider the distance between the rescaled modified cumulative distribution function and empirical distribution function obtained for vector $U$. The main issue during estimation of parameters corresponding to distribution of constant time periods comes from the fact that the exact waiting time is unknown and usually comes from continuous distribution. For example, if we observe that a character of the process has changed after 3 units of time, it is not known at which point of
time the change actually happened, the correct value lies in the interval (2, 4). Due to this fact the modified version of cumulative distribution function was introduced. In \cite{rafal} it is proved that the distance between empirical cumulative distribution function (CDF) of waiting times from given distribution and rescaled modified distribution function is smallest according to the distance between empirical CDF and the  theoretical CDF of given distribution. First we introduce the modified and rescaled modified CDF.   The modified cumulative distribution function of given distribution with CDF $F(\cdot)$ can be expressed as \cite{rafal}
\begin{eqnarray}
\tilde{F}(n)=\int_{n}^{n+1}F(x)dx.
\end{eqnarray} 
The rescaled modified cumulative distribution function is defined in the following way
\begin{eqnarray*}
&G(0) = 0 \\
&G(n) = \frac{\tilde{F}(n)-\tilde{F}(0)}{1-\tilde{F}(0)} ~~ \mbox{for $n \geq 1$.} 
\label{finalCDF}
\end{eqnarray*}
For each class of cumulative distribution functions  we can find parameters which minimize the distance between empirical CDF and rescaled modified cumulative distribution function which satisfies the following condition
\begin{eqnarray}
D(G_{\theta_0}, \hat{F}) = \inf_{\theta \in \Theta}{D(G_{\theta}, \hat{F})},
\label{minDistance}
\end{eqnarray}
where $D$ is one of the introduced distances: $KS$, $CvM$ and $AD$, $\hat{F}$ is empirical CDF, $G$ is a rescaled cumulative distribution function defined in (\ref{finalCDF}) and $\Theta$ is the set of parameters of a certain class of distribution functions, in our case the $\nu$ parameter from gamma distribution.
The Hurst parameter $H$ is estimated on the basis of the vector ${B_H}$. The parameter $H=d+1/2$ can be obtained by using the sample MSD, that for FBM asymptotically behaves in distribution as power law, see equation  (\ref{power}). By comparing the sample MSD and the power function $\tau^{2d+1}$ we can estimate parameter $d$ (and thus $H$) by using the least squares method. \\
Similar to previous case, we simulate $100$ trajectories of the process $Y_2(t)$ and for each of them we estimate the unknown parameters. The $\nu$ parameter is estimated by using three mentioned distances, namely KS, CvM and AD. We consider here two sets of parameters: $H=0.3$ and $\nu=1$ and $H=0.7$ and $\nu=1$. In Figs. \ref{fig6} and \ref{fig7} we present the obtained results. It can be seen the estimated values coincide with the theoretical ones which confirms the effectiveness of the estimation procedure.

\section{Conclusions}
In this paper we have analyzed two processes based on the FBM. The first one is a time-changed FBM by gamma process while the second one is the FBM delayed by inverse to gamma process. Those processes can be universal models for modeling data with special behavior. The first analyzed process, known in the literature as the FLM, can be useful for modeling data with LRD which is not Gaussian, however possess properties similar to Gaussian processes, like existence of all moments. The second considered process can be used for data with LRD and visible constant time periods characteristic for processes delayed by inverse subordinators. We have compared the main properties of considered time-changed processes and explained how to simulate them. Moreover we have described also the estimation procedures for parameters of both processes and checked their effectiveness by using simulated data. We hope the introduced estimation schemes can be applied to different real data which have similar properties as the analyzed systems.\\  The considered processes can be generalized by replacing the gamma subordinator and its inverse  by other subordinators having their origins in the wide class of infinitely divisible processes.

\section*{Acknowledgments}
AW would like to acknowledge a support of  NCN Maestro Grant No. 2012/06/A/ST1/00258.

\section*{References}


\begin{thebibliography}{00}
\bibitem{Barn-Nic-Shep} O. Barndorff-Nielsen, E. Nicolato, N.Shephard,  Quantitative Finance 2, 11, 2002.
\bibitem{wyl1}A. Wy{\l}oma{\'n}ska, Physica A 391 (22), 5685, 2012.
\bibitem{boch}S. Bochner, Proc. Nat. Acad. Sci USA 35, 368 1949.
\bibitem{boch2}S. Bochner, Harmonic Analysis and the Theory of Probability, Unif. California Press, 1955.
\bibitem{sato}K. -I. Sato, L\'evy Processes and Infinitely Divisible Distributions, Cambridge University Press, 1999.
\bibitem{fin1}P. Clark, Econometrica 41, 135, 1973.
\bibitem{fin2}X. Gabaix, P. Gopikrishnan, V. Plerou, H.E. Stanley, Nature 423, 267, 2003.
\bibitem{fin3}P. Ch. Ivanov, A Yuen, B. Podobnik, Y. Lee, Phys. Rev. E 69, 056107, 2004.
\bibitem{fin4}B. Podobnik, D. Wang,  H. E. Stanley, Quantitative Finance 12, 559, 2012.
\bibitem{ding-granger} Z. Ding, C. W. J. Granger, R. F. Engle,  J. Empirical Finance 1, 83, 1993.
\bibitem{pagan} A. Pagan,  J. Empirical Finance 3, 15, 1996.
\bibitem{fiz1}M. G. Nezhadhaghighi, M. A. Rajabpour, S. Rouhani, Phys. Rev. E 84, 011134, 2011.
\bibitem{fiz2}R. Failla, P. Grigolini, M. Ignaccolo, A. Schwettmann, Phys. Rev. E 70, 010101(R), 2004.
\bibitem{fiz3}A. Stanislavsky, K. Weron, Ann. Phys. 323(3), 643, 2008.
\bibitem{fiz4} B. Dybiec, E. Gudowska-Nowak, Chaos 20(4),  043129, 2010.
\bibitem{ec} H. Scher, G. Margolin, R. Metzler, J. Klafter, Geophys. Res. Lett. 29, 1061, 2002.
\bibitem{dok-opp-taqqu} P. Doukhan, G. Oppenheim, M. S. Taqqu (Eds.), Theory and applications of long-range dependence. Birkh\'oauser Boston, Inc., Boston, 2003.
\bibitem{bi} I. Golding, E. C. Cox, Phys. Rev. Lett. 96, 098102, 2006.
\bibitem{k1} B. B. Mandelbrot, J. R. Wallis, Water Resour. Res. 4, 909, 1968. 
\bibitem{NessMandelbrot} B. B. Mandelbrot, J. W. Van Ness, SIAM Rev. 10(4), 422, 1968.
\bibitem{Metzler_fractional} E. Lutz, Phys. Rev. E 64, 051106, 2001.  
\bibitem{k2} A. W. Lou, Econometrica, 59, 1279, 1991. 
\bibitem{k4} B. B. Mandelbrot, The Fractal Geometry of Nature, San Francisco, Freeman, 1982. 
\bibitem{k5} J. Beran, Statistics for Long-Memory Processes, New York, Chapman $\&$ Hall, 1994.
\bibitem{k7} M. Bertacca, F. Berizzi, E. Mese, IEEE Trans. Geosci. Remote Sens., 43, 2484, 2002. 
\bibitem{k8}  A. Stanislavsky, K. Burnecki, M. Magdziarz, A. Weron, K. Weron,  Astrophys. J. 693, 1877, 2009. 
\bibitem{k9}  D. Horvatic, H. E. Stanley, B. Podobnik,  Europhys. Lett. 94, 18007, 2011. 
\bibitem{w47} K. Burnecki, A. Weron, J. Stat. Mech. P10036, 2014. 
\bibitem{heyde-leonenko} C. C. Heyde, N. N. Leonenko, Adv. Appl. Probab. 37, 342, 2005.
\bibitem{meer}T. J. Kozubowski, M. M. Meerschaert, K. Podgorski,  Adv. Appl. Prob. 38, 451, 2006.
\bibitem{ar30}E. W. Montroll, G. H. Weiss, J. Math. Phys. 6, 167, 1965.
\bibitem{ar31}H. Scher, E. Montroll, Phys. Rev. B 12, 2455, 1975.
\bibitem{ar32}M. Magdziarz, A. Weron, K. Weron, Phys. Rev. E 75, 016708, 2007.
\bibitem{ar33}S. Orze{\l}, A. Wy{\l}oma{\'n}ska, J. Stat. Phys. 143, 447, 2011.
\bibitem{ar34}M. Magdziarz, Stoch. Proc. Appl. 119, 3238, 2008.
\bibitem{ar35}A. Stanislavsky, Phys. Scr. 67, 265, 2003.
\bibitem{ar36}A. Stanislavsky, K. Weron, A. Weron, Phys. Rev. E 78, 051106, 2008.
\bibitem{ar38}J. Gajda, M. Magdziarz, Phys. Rev. E 84, 021137, 2011.
\bibitem{ar39}J. Janczura, A. Wy{\l}oma{\'n}ska, Acta Phys. Polon. B 43(5), 1001, 2012.
\bibitem{ar40}T. R. Hurd, A. Kuznetsov, J. Appl. Probab. 46, 181, 2009.
\bibitem{ar41}M. Magdziarz, J. Stat. Phys. 135 (2009) 763.
\bibitem{ar42}A. Piryatinska, A. I. Saichev, W. A. Woyczynski, Physica A 349, 375, 2004.
\bibitem{rafal}R. Po{\l}ocza{\,n}ski, A. Wy{\l}oma{\'n}ska, J. Gajda,  M. Maciejewska, A. Szczurek, Modified cumulative distribution function in application to waiting times analysis in CTRW scenario, arXiv:1604.02653, 2016.
\bibitem{burnwer}K. Burnecki and A. Weron,  Phys. Rev. E 82, 021130, 2010.
\bibitem{sierra}A. Szczurek, M. Maciejewska, R. Po{\l}ocza{\'n}ski, M. Teuerle, A. Wy{\l}oma{\'n}s, Stoch. Environ. Res. Risk. Assess. 29(8), 2193, 2015.
\bibitem{b121}A. Kolmogorov, Wienersche Spiralen und eigene eigene andere interessante Kurven in Hilberschen Raum, C.R. (Doklady) Acad SciURSS (N.S.) 26, 115, 1940.
\bibitem{marek}M. Teuerle, A. Wy{\l}oma{\'n}ska, G. Sikora, J. Stat. Mech. P05016, 2013.
\bibitem{codi}A. Wy{\l}oma{\'n}ska, A. Chechkin, I. M. Sokolov, J. Gajda, Physica A 421, 412, 2015.
\bibitem{gaj2}J. Gajda, A. Wy{\l}oma{\'n}ska, Physica A 405, 104, 2014.
\bibitem{klafter}M. Magdziarz, A. Weron, J. Klafter,  Phys. Rev. Let. 101, 210601, 2008.
\bibitem {applebaum} D. Applebaum,  L\'evy Processes and Stochastic Calculus. 2nd ed., Cambridge University Press, Cambridge, U.K., 2009.
\bibitem{meerschaert-scheffler} M. M. Meerschaert, H. Scheffler, Stochastic Process. Appl. 118, 1606, 2008.
\bibitem{schiff} J. L. Schiff, The Laplace Transform: Theory and Applications. Springer-Verlag, New York, 1999.
\bibitem{abramowitz} M. Abramowitz, I. A. Stegun (Eds.), Handbook of Mathematical Functions with Formulas, Graphs and Mathematical Tables. Dover, New York, 1992.
\bibitem{kumar} A. Kumar, P. Vellaisamy, Stat. Probab. Lett. 103, 134, 2015.
\bibitem{bertoin} J. Bertoin, L\'evy Processes, Cambridge University Press, Cambridge, 1996.
\bibitem {mijena} J. B.  Mijena, Correlation structure of time-changed fractional Brownian
motion, arXiv:1408.4502, 2014.
\bibitem{cont} R. Cont, P. Tankov, Financial Modeling with Jump Processes. Chapman \& Hall CRC Press, Boca Raton, 2004.
\bibitem{m60}S. Stoev, M. Taqqu, Fractals 12, 95, 2004.
\bibitem{boxplot}Y. Benjamini,  Amer Statist. 42, 257, 1988.
\bibitem{krzysiek}K. Burnecki, A. Wy{\l}oma{\'n}ska, A. Chechkin, PLoS ONE 10(12), 2015.
\bibitem{orzel_ja} S. Orze{\l}, A. Wy{\l}oma{\'n}ska, J. Stat. Phys. 143(3), 447, 2011.
\bibitem{janusz_ja} J. Gajda, A. Wy{\l}oma{\'n}ska, J. Stat. Phys. 148, 296, 2012.

\end{thebibliography}
\end{document}